\documentstyle[prd,aps,epsfig]{revtex}
\newcommand{\be}{\begin{equation}}
\newcommand{\ee}{\end{equation}}
\newcommand{\ba}{\begin{eqnarray}}
\newcommand{\ea}{\end{eqnarray}}

\newcommand{\ov}{\overline}

\newcommand{\aand}{\;\;\;\mbox{and}\;\;\;}
\newcommand{\pa}{\partial}
\def\sl#1{\rlap{\hbox{$\mskip 1 mu /$}}#1}
\def\Sl#1{\rlap{\hbox{$\mskip 3 mu /$}}#1}
\def\SL#1{\rlap{\hbox{$\mskip 4.5 mu /$}}#1}

\newcommand{\pari}{\stackrel{{P}}\longrightarrow}

\begin{document}

\draft
%\twocolumn
\title{Electronic bound states in parity-preserving QED$_{3}$ applied to
high-T$_{\rm c}$ cuprate superconductors}
\author{
H.R. Christiansen$^{b}$\thanks{\tt hugo@cbpf.br},
O.M. Del Cima$^{a}$\thanks{\tt delcima@gft.ucp.br},
M.M. Ferreira Jr.$^{b,c}$\thanks{\tt manojr@cbpf.br} and
J.A. Helay\"el-Neto$^{a,b}$\thanks{\tt helayel@gft.ucp.br}     }
\address{
$^a${\it Grupo de F\'\i sica Te\'orica (GFT), \\
Universidade Cat\'olica de Petr\'opolis (UCP), \\
Rua Bar\~ao do Amazonas 124 - 25685-070 - Petr\'opolis - RJ - Brazil.}\\
$^b${\it Centro Brasileiro de Pesquisas F\'\i sicas (CBPF),\\
Coordena\c c\~ao de Teoria de Campos e Part\'\i culas (CCP), \\
Rua Dr. Xavier Sigaud 150 - 22290-180 - Rio de Janeiro - RJ - Brazil.}\\
$^c${\it Universidade Federal do Maranh\~ao (UFMA), \\
Departamento de F\'\i sica, \\
Campus Universit\'ario do Bacanga - 65085-580 - S\~ao Luiz - MA - Brazil.}}

\date{\today}
\maketitle

%==============================================================
\begin{abstract}
We consider a parity-preserving QED$_3$ model with
spontaneous breaking of the gauge symmetry as a framework
for the evaluation of the electron-electron interaction potential
underlying high-T$_{\rm c}$ superconductivity. The fact that the resulting
potential, $-C_sK_{0}(Mr)$, is non-confining and ``weak''
(in the sense of Kato) strongly suggests the mechanism
of pair-condensation. This potential, compatible with an $s$-wave order parameter,
is then applied to the Schr\"odinger equation for the sake
of numerical calculations, thereby enforcing the
existence of bound states. The results worked out by means of our
theoretical framework are checked by considering a number of
phenomenological data extracted from different copper oxide superconductors.
The agreement may motivate a deeper analysis of our model viewing an application
to quasi-planar cuprate superconductors. The data analyzed here
suggest an energy scale of $1$-$10$meV for the breaking of the $U(1)$-symmetry.
\end{abstract}
\pacs{PACS numbers: 11.10.Kk 11.15.Ex 74.20.-z 74.72.-h
\hspace{4,05cm}ICEN-PS-01/17}
%==============================================================
\section{Introduction}
Planar QED (QED$_3$) has shown to be an appropriate
theoretical framework for discussing issues of contemporary physics,
particularly in connection with Condensed Matter Physics. In the latest years,
the raising interest in applications of this theory to high-T$_{\rm c}$
superconductivity and quantum
Hall effect \cite{Q-hall} has motivated an enormous production of works in this
subject. The  relation between QED$_3$ and
superconductivity, phenomenon discovered in 1986 \cite{Bednorz},
 can be traced back to 1987, when Anderson \cite{Anderson} suggested
that in some copper oxide superconductors (based on La$_2$CuO$_4$)  the
hypothesized resonant-valence-bond state or quantum-spin-liquid state of
Mott (a kind of insulator) could migrate to a superconducting state
through a doping process. Soon after, in 1988, Laughlin \cite{Laughlin1}
argued that the excitations of the Anderson resonating-valence-bond model
behaved like fractional quantum Hall states (anyons), presenting consequently
a fractional or anyonic statistics. Despite the initial success of this model,
several difficulties with this idea arose. The main problem concerns the
necessity of a massless scalar mode in the spectrum which occurs only
when the bare Chern-Simons term
cancels with the term generated by one-loop radiative corrections.
This cancellation occurs exactly at zero temperature,
but does not take place at finite temperature.  In this way, one can assert
that the anyonic model behaves like a superconductor only at zero
temperature \cite{Lykken-Dorey}.

At the same time that the anyonic model was developed, a new approach based upon
the QED$_3$ theoretical framework \cite{Kogan} began to be adopted to
explain the formation of electron-electron bound states, provided that the
high-T$_{\rm c}$ superconductors had quasi-planar structure. In the domain of
QED$_3$, there arises the necessity of providing the gauge field with a mass
in order to circumvent the appearance of a confining potential associated to
the long-range Coulomb interaction. The Maxwell-Chern-Simons model is then
adopted so as to generate (topological) mass for the photon, leading to a
finite range interaction, to which a binding potential is associated
instead of a confining one. In the framework of a Maxwell-Chern-Simons theory, numerical
evaluation of electron-electron bound states were first addressed to in
Ref.\cite{Girotti}, but the assumptions and results of the latter induced
some controversy \cite{Hagen,Dobroliubov}.  Other authors
\cite{Groshev-Georgelin}, working in this same context, have also obtained
bound states, corresponding to the situation where the magnetic-dipole
interaction between the electrons is large enough to overcome the
Coulombian repulsion. In this case, however, the attractive interaction
only appears when the topological mass of the gauge field is larger
than the electron mass ($\kappa>m$). This condition prevents
the application of this model to Condensed Matter systems, where one must have
$\kappa\ll m$ due to the order of magnitude ($\sim{\rm meV}$)
of the usual relevant excitations.
An attempt to bypass this difficulty consists in considering a Maxwell-Chern-Simons
model minimally coupled to fermions and bosons with spontaneous breaking of a local
$U(1)$-symmetry as a generating mechanism for the photon mass \cite{MCS_PROCA}, whose
results show the possibility of obtaining bound states whenever the attractive
Higgs interaction dominates over the gauge boson interchange.
This issue is now under investigation \cite{boundMCS}, which would be
suitable to apply such a model to the cases where there is
an evidence of a parity-breaking superconductivity quasi-planar phenomenon
\cite{Parity-breaking}. Very recently, has been
proposed in Ref.\cite{franz}, an anisotropic $U(1)\times U(1)$ QED$_3$ model
coupled to both the ``Berry'' and ``Doppler'' gauge fields,
by arguing that the pseudogap regime in cuprates could be modeled as a phase
disordered $d$-wave superconductor.

In the present work we consider a parity-preserving
QED$_3$ model with spontaneous breaking of the local $U(1)$-symmetry
accomplished by a sixth-power potential \cite{N.Cimento}. Our aim here is to carry
out numerical calculations in searching for electron-electron bound states,
in such a theoretical framework. The breaking mechanism of $U(1)$-symmetry
gives rise to a Higgs-type boson and a massive photon avoiding the appearance of a
confining logarithmic potential (characteristic of massless interactions in
three space-time dimensions).
Hence, the Higgs mechanism has the relevant role of contributing to the
electron-electron binding while yielding a non-confining potential.

Thereafter, the consideration of the M{\o}ller scattering mediated
by both the vector and scalar bosons results in the establishment of an
attractive electron-electron
potential, independent of the spin polarization state. The potential stemming
from the M{\o}ller scattering corresponds to a modified Bessel function of
zeroth order, $-C_sK_0(Mr)$, that besides being non-confining, assures the
semi-boundedness of the system (the so-called weak Kato condition).
Once we have proven that the $K_0$-type potential
satisfies the necessary conditions to allow the existence of
bound states, a numerical calculation of the ground state energy of the Schr\"odinger
equation is carried out. Incidentally, by virtue of the radial symmetry of the potential,
we are bound to only consider the $s$-wave solutions. An application of these
numerical calculations to high-T$_{\rm c}$ superconductivity is then implemented by
fitting some phenomenological data available for the following cuprate
superconducting materials: YBa$_2$Cu$_3$O$_7$, Tl$_2$Ba$_2$CaCu$_2$O$_{10}$,
Bi$_2$Sr$_2$CaCu$_2$O$_8$ and HgBa$_2$Ca$_2$Cu$_3$O$_8$. Our procedure reveals to
be successful in the sense that it is always possible to fit the energy gap
energy and the correlation length of the samples through the indication of
a specific scalar vacuum expectation value (v.e.v.).

The outline of this paper is the following. In Section \ref{sec2}, we present the model.
Next, in Section \ref{sec3}, we address the relevant
Schr\"odinger equation taking into account the properties of
the condensate wave-function. Some aspects of the trial
function are discussed so that it turns to be suitable to the variational method.
In Section \ref{sec4}, we digress on some aspects of copper oxide superconductors, as the
order parameter, the pairing mechanism and the effective coupling constant.
Section \ref{sec5} is devoted to the identification of free pure theoretical
parameters with phenomenological ones, and finally, in Section \ref{sec6}, we
perform a numerical calculation of the energy gap and correlation length for
four cuprate high-T$_{\rm c}$ superconducting samples.

%%%%%%%%%%%%%%%%%%%%%%%%% SECTION 2 %%%%%%%%%%%%%%%%%%%%%%%%%%%%%%%%%%%%%%
\section{Brief survey on the parity-preserving QED$_{3}$}\label{sec2}
%%%%%%%%%%%%%%%%%%%%%%%%% SECTION 2 %%%%%%%%%%%%%%%%%%%%%%%%%%%%%%%%%%%%%%
The action for the parity-preserving QED$_{3}$\footnote{The metric is
given by $\eta_{\mu\nu}=(+,-,-)$; $\mu$,$\nu$=(0,1,2) and the $\gamma$-matrices
are taken as $\gamma^\mu=(\sigma_x,i\sigma_y,-i\sigma_z)$.} with spontaneous symmetry
breaking of a local $U(1)$-symmetry is given by \cite{phdthesis,N.Cimento,Del_cima}:
\ba
S_{{\rm QED}}&=&\int{d^3 x} \biggl\{ -{1\over4}F^{\mu\nu}F_{\mu\nu}
+ i {\ov\psi _+} {\SL{D}} {\psi}_+ + i
{\ov\psi _-} {\SL{D}} {\psi}_- - m_e (\ov\psi_+\psi_+ -
\ov\psi_-\psi_-) - y (\ov\psi_+\psi_+ -
\ov\psi_-\psi_-)\varphi^*\varphi + \nonumber\\
&+& D^\mu\varphi^* D_\mu\varphi - V(\varphi^*\varphi)\biggr\}~,
\label{action1}
\ea
with the potential $V(\varphi^*\varphi)$ taken as
\be
V(\varphi^*\varphi)=\mu^2\varphi^*\varphi + {\zeta\over2}(\varphi^*\varphi)^2 +
{\lambda\over3}(\varphi^*\varphi)^3~,
\label{potential}
\ee
where the mass dimensions of the parameters
$\mu$, $\zeta$, $\lambda$ and $y$ are respectively ${1}$, ${1}$, ${0}$ and ${0}$.
The sixth-power potential, $V(\varphi^*\varphi)$, is the responsible
for breaking the electromagnetic $U(1)$-symmetry.

The covariant derivatives are defined as follows:
\be
{\SL{D}}\psi_{\pm}\equiv(\sl{\pa} + ie \Sl{A})\psi_{\pm}
\aand
D_{\mu}\varphi\equiv(\pa_{\mu} + ie A_{\mu})\varphi~,
\label{covder}
\ee
where $e$ is a coupling constant with dimension of
(mass)$^{1\over2}$. In the
action (\ref{action1}), $F_{\mu\nu}$ is the usual field
strength for $A_\mu$, $\psi_+$ and $\psi_-$ are two kinds
of fermions (the $\pm$ subscripts refer to their spin sign
\cite{phdthesis,N.Cimento,Binegar})
and $\varphi$ is a complex scalar. The $U(1)$-symmetry gauged by $A_\mu$ is
interpreted as the electromagnetic one, so that $A_\mu$ is meant to describe
the photon.

The action given by Eq.(\ref{action1}) is invariant under the discrete
symmetry, $P$, whose action is fixed below:
\ba
x_\mu &\pari& x_\mu^P=(x_0,-x_1,x_2), \nonumber\\
\psi_{\pm} &\pari& \psi_{\pm}^P=-i\gamma^1\psi_{\mp}
~,~\ov\psi_{\pm} ~\pari ~\ov\psi_{\pm}^P=i\ov\psi_{\mp}\gamma^1~, \nonumber\\
A_\mu &\pari& A_\mu^P=(A_0,-A_1,A_2)~, \nonumber\\
\varphi &\pari& \varphi^P=\varphi~.
\ea

Analyzing the potential (\ref{potential}), and
imposing that it is bounded from
below and yields only stable vacua (metastability is
ruled out), the following
conditions on the parameters $\mu$, $\zeta$, $\lambda$ must be set:
\be
\lambda>0 ~,~ \zeta<0 \aand \mu^2 \leq {3\over 16}
{\zeta^2\over \lambda}~.
\label{cond}
\ee
We denote ${\langle}\varphi{\rangle}=v$ and the v.e.v. for the
$\varphi^*\varphi$-product, $v^2$, is chosen as
\be
{\langle}\varphi^*\varphi{\rangle}=v^2=-{\zeta \over 2\lambda}+
\left[ \biggl({\zeta \over
2\lambda}\biggr)^2 - {\mu^2\over \lambda} \right]^{1\over 2}~, \label{vac}
\ee
the condition for minimum being read as
\be
\mu^2+{\zeta}v^2+{\lambda}v^4=0~. \label{mincond}
\ee

The complex scalar, $\varphi$, is parametrized by
\be
\varphi=v+H+i\theta~, \label{para}
\ee
where $\theta$ is the would-be Goldstone boson and $H$
is the Higgs scalar, both
with vanishing v.e.v.'s.

By replacing the parametrization (\ref{para}) for
the complex scalar, $\varphi$,
into the action (\ref{action1}), the following free
action comes out:
\ba
S_{{\rm QED}}^{\rm free}&=&
\int{d^3 x}\biggl\{-{1\over4}
F^{\mu\nu}F_{\mu\nu} + {1\over2} M^2_A A^\mu A_\mu
+  {\ov\psi _+} (i{\sl{\pa} - m_{\rm eff}}) {\psi}_+ +
{\ov\psi _-} (i{\sl{\pa} + m_{\rm eff}}) {\psi}_- + \nonumber\\
&+& \pa^\mu H \pa_\mu H - M^2_H H^2 + {\pa^\mu}\theta {\pa_\mu}\theta +
2ve{A^\mu} {\pa_\mu}\theta \biggr\}~,\label{action2}
\ea
where the parameters $M^2_A$, $m_{\rm eff}$ and $M^2_H$ are
given by
\be
M^2_A=2v^2e^2~,~m_{\rm eff}=m_e+yv^2 \aand
M^2_H=2v^2(\zeta+2 \lambda v^2)~.\label{masses}
\ee
The conditions (\ref{cond}) and (\ref{mincond})
imply the following lower-bound
(see Eq.(\ref{masses})) for the Higgs mass:
\be
M^2_H \geq {3\over 4} {\zeta^2\over \lambda}~.
\label{lowbound}
\ee
Therefore, a {\it massless} Higgs is out of the model
we consider here, it would be present in the spectrum if
$\mu^2>3\zeta^2/16\lambda$. However, in such a situation,
the minima realizing the
spontaneous symmetry breaking would not be absolute
ones, corresponding
therefore to metastable ground states, that we avoid
here. One-particle states
would decay with a short decay-rate if compared to
an absolute minimum ground
state.

In order to preserve the manifest renormalizability
of the model, the 't Hooft gauge {\cite{thooftg}} is adopted:
\be
{\hat S}_{R_{\xi}}^{\rm gf}=\int{d^3 x}\biggl\{-{1\over 2\xi}
\left(\pa^\mu A_\mu -{\sqrt{2}}\xi M_A \theta \right)^2 \biggr\}~,
\label{gf}
\ee
where $\xi$ is a dimensionless gauge parameter.

By replacing the parametrization (\ref{para})
into the action (\ref{action1}),
and adding up the 't Hooft gauge (\ref{gf}), it
can be directly found the
following complete parity-preserving action:
\ba
{S}_{{\rm QED}}^{\rm SSB}&=&\int{d^3 x}\biggl\{-{1\over4}
F^{\mu\nu}F_{\mu\nu} + {1\over2} M^2_A A^\mu A_\mu
+  {\ov\psi _+} (i{\sl{\pa} - m_{\rm eff}}) {\psi}_+ +
{\ov\psi _-} (i{\sl{\pa} + m_{\rm eff}}) {\psi}_- + \nonumber\\
&+& \pa^\mu H \pa_\mu H - M^2_H H^2 +
{\pa^\mu}\theta {\pa_\mu}\theta - M^2_\theta \theta^2 -{1\over 2\xi}
(\pa^\mu A_\mu)^2 + \nonumber \\
&-& e {\ov\psi _+}{\Sl{A}}{\psi}_+ - e {\ov\psi _-}
{\Sl{A}}{\psi}_-
-  y (\ov\psi_+\psi_+ - \ov\psi_-\psi_-) (2vH + H^2 +\theta^2) + \nonumber \\
&+& e^2 A^\mu A_\mu (2vH + H^2 +\theta^2) + 2eA^\mu(H{\pa_\mu}\theta - \theta{\pa_\mu}H)+ %%@
\nonumber \\
&-& c_3 H^3 - c_4 H^4 - c_5 H^5 - c_6 H^6 - c_7\theta^4 - c_8\theta^6 -
c_9H\theta^2 - c_{10}H^2\theta^2 + \nonumber \\
&-& c_{11}H^3\theta^2 - c_{12}H^4\theta^2 - c_{13}H\theta^4 - c_{14}H^2\theta^4 %%@
\biggr\}~,\label{action3}
\ea
where the constants $M^2_\theta$, $c_3$, $c_4$, $c_5$, $c_6$, $c_7$, $c_8$,
$c_9$, $c_{10}$, $c_{11}$, $c_{12}$, $c_{13}$ and $c_{14}$ are defined by
\ba
&&M^2_\theta=\xi M^2_A~,~ c_3=2v(\zeta+{10\over3}\lambda v^2)
~,~c_4={\zeta\over 2}+5\lambda v^2~,~c_5=2\lambda v~,\nonumber\\
&&c_6={\lambda\over 3}~,~c_7={\zeta\over2}+\lambda v^2~,~c_8={\lambda\over 3}
~,~c_9=2v(\zeta + 2\lambda v^2)~,\nonumber\\
&&c_{10}=\zeta + 6\lambda v^2~,~c_{11}=4\lambda v~,~c_{12}=\lambda
~,~c_{13}=2\lambda v \aand c_{14}=\lambda~.
\label{chiggs}
\ea

Working in the 't Hooft gauge, one evaluates, in the non-relativistic limit,
the M{\o}ller scattering potential mediated by the Higgs quasi-particle
and the photon in the center-of-mass frame. In the Born approximation the
potential is nothing but the Fourier transform of the total M{\o}ller
scattering amplitude, yielding, therefore, to the following the electron-electron
scattering potential:
\be
V(r)=-\frac{1}{2\pi}\left[2y^2v^2 K_0(M_Hr)-
e^2 K_0(M_Ar)\right]~.\label{potential1}
\ee
However, the potential thus obtained is attractive provided the attraction caused by
the Higgs quasi-particle mediation overcomes the repulsion resulted from the photon
mediation. In fact, as one shall present later, the quantity, $C_s=2y^2v^2-e^2$,
varies from $3,01$ to $7,22$meV for the copper oxides superconductors analyzed here.

%%%%%%%%%%%%%%%%%%%%%%%%% SECTION 3 %%%%%%%%%%%%%%%%%%%%%%%%%%%%%%%%%%%%%%
\section{The composite wave-function and the Schr\"odinger equation}\label{sec3}
%%%%%%%%%%%%%%%%%%%%%%%%% SECTION 3 %%%%%%%%%%%%%%%%%%%%%%%%%%%%%%%%%%%%%%
Before studying the Schr\"odinger equation, it is instructive to analyze
the behavior of the total wave-function $(\Psi)$ of a two-electron
system in light of the Pauli exclusion principle. By exchanging both
fermions, one knows that $\Psi$ must be antisymmetric with
respect to a permutation between them
\be
\Psi({\bf \rho}_{1},s_{1,}{\bf \rho}_{2},s_{2})=
-\Psi({\bf \rho}_{2},s_{2,}{\bf \rho}_{1},s_{1}).
\ee

Assuming that no significant spin-orbit interaction takes place,
this function can be written in terms of the following three
independent functions:
\be
\Psi(\rho_{1},s_{1,}\rho_{2},s_{2})=\psi({\bf R})
\varphi({\bf r}_{1},{\bf r}_{2})\chi\left(s_{1},s_{2}\right)~,
\ee
which represent, respectively, the center-of-mass wave function,
the relative one, and the spin wave function (${\bf R}$ and $s$ being the center of mass
and spin coordinates respectively, while ${\bf r}_1$ and ${\bf r}_2$ are the
electrons coordinates relative to ${\bf R}$).

Each of these functions contain information on the mechanism
underlying superconductivity. Flux quantization results from the boundary
conditions on $\psi$, and from this it can be deduced that the
charge carriers are pairs of particles \cite{little}. The other two
functions tell us about other features of the condensates. For instance,
the radial component of $\varphi$ has information on the spatial extent
of the pairs, and the rest of the factors determine whether they are
in an $s$, $p$ or $d$ state, or even whether the system is in a
singlet or triplet spin-state.

For the spin singlet ($S=0$), the spin-function, $\chi$, is
antisymmetric, while for the spin triplet ($S=1$) it is symmetric.
Consequently, the space-function $\varphi(r)$ associated with a spin triplet
must be even, and the one associated with a spin singlet must be odd:
\ba
\Psi^{S=1}&=&\varphi_{\rm odd}({\bf r}_{1},{\bf r}_{2})
\chi_{\rm even}^{S=1}(s_{1},s_{2})~,  \label{antisym1} \nonumber\\
\Psi^{S=0}&=&\varphi_{\rm even}({\bf r}_{1},{\bf r}_{2})
\chi_{\rm odd}^{S=0}(s_{1},s_{2})~.   \label{antisym2}
\ea

Thus, by only considering the Pauli exclusion principle, one concludes
that the total wave-function will be composed by an even angular momentum
state ($s$-wave, $d$-wave) and a $s$-spin state, or by an
odd angular momentum state ($p$-wave, $f$-wave) and a $p$-spin state.

Consider now the planar Schr\"odinger equation for the relative
wave-function, $\varphi(r)$, representing an electron-electron system,
with relative radial coordinate $r$:
\be
\frac{\partial^2\varphi(r)}{\partial r^2} + \frac{1}{r}\frac{\partial
\varphi(r)}{\partial r} - \frac{l^{2}}{r^2}\varphi(r) +
2\mu_{\rm eff}[E-v(r)]\varphi(r)=0~, \label{diff1}
\ee
where $V(r)$ represents the interaction potential between the two electrons,
Eq.(\ref{potential1}), and $\mu_{\rm eff}$ the effective reduced mass of the system
\be
\mu_{\rm eff}=\frac{1}{2}(m_{e}+yv^{2})~.
\ee

By means of the following transformation:
\be
\varphi(r)=\frac{1}{\sqrt{r}}~g(r)~,
\ee
one has
\be
\frac{\partial^2g(r)}{\partial r^2} - \frac{l^2-\frac{1}{4}}{r^2}g(r) +
2\mu_{\rm eff}[E-V(r)]g(r)=0~.\label{diff2}
\ee
Looking at this equation, it is easy to identify the effective potential
for the electron-electron system as:
\be
V_{\rm eff}(r)=\frac{l^2-\frac{1}{4}}{2\mu_{\rm eff}r^2}-C_sK_{0}(M_{H}r)~,
~C_{s}=\frac{1}{2\pi}(2y^2v^2-e^2)~, \label{cs}
\ee
where, for the sake of simplicity, we considered equal masses for the scalar and vector
excitations in order to check the possibility of bound states in the
model. However, we should stress that the assumption of equal masses might lead
to conflicts with phenomenological data. Later, as we shall discuss, this will not
be the case; in any case, it is a sensible attitude to reassess our results by taking
$M_H\neq M_A$.

%%%%%%%%%%%%%%%%%%%%%%%%% subsection 3.1 %%%%%%%%%%%%%%%%%%%%%%%%%%%%%%%%
\subsection{The choice of the trial function}
%%%%%%%%%%%%%%%%%%%%%%%%% subsection 3.1 %%%%%%%%%%%%%%%%%%%%%%%%%%%%%%%%
The variational method is used for the approximate determination of the
ground state energy level, and consists in determining the
wave-function $\varphi(r)$ that provides the largest (absolute) binding energy
value. This method is applied mainly in
situations where the wave function for the system is unknown, depending on the
choice of a trial function. The closer the trial function is to the true
solution of the system, the better the energy value numerically obtained will be.
The definition of a trial function must observe some conditions, such as the
asymptotic behavior at infinity, the analysis of its free version and its
behavior at the origin. For a zero angular momentum ($l=0$) state,
Eq.(\ref{diff2}) becomes
\be
\biggl\{\frac{\partial^{2}}{\partial r^2} + \frac{1}{4r^2} +
2\mu_{\rm eff}[E+C_{s}K_{0}(M_{H}r)]\biggr\}g(r)=0~. \label{diff3}
\ee

The free version, $V(r)=0$, of the last equation is given merely by
\be
\left[\frac{\partial^{2}}{\partial r^2} + \frac{1}{4r^2} + k^2\right]u(r)=0~,
\label{diff4}
\ee
whose solution is
\be
u(r)=B_{1}\sqrt{r}J_{0}(kr)+B_{2}\sqrt{r}Y_{0}(kr)~, \label{asymptotic}
\ee
where $B_{1}$ and $B_{2}$ are arbitrary constants and $k=\sqrt{2\mu_{\rm eff}E}$.
In the limit $r\rightarrow 0$, Eq.(\ref{asymptotic}) goes simply as
\be
u(r)\longrightarrow\sqrt{r}+\lambda\sqrt{r}\ln(r)~. \label{asymptotic2}
\ee

Since the second term in Eq.(\ref{diff4}) behaves like an attractive
potential, $-1/4r^2$, this implies the possibility of
obtaining a bound state ($E<0$) even for $V(r)=0$
\cite{Tichmarsh,Chadan}. This is not physically acceptable,
leading to a restriction on the needed self-adjoint
extension of the differential operator $-d^2/dr^2-1/4r^2$. Among the infinite
number of self-adjoint extensions of this
symmetric operator, the only physical choice corresponds to the
Friedrichs extension ($B_2=0$), which behaves like $\sqrt r$
at the origin.
The choice of the Friedrich extension thus circumvents the non-physical
possibility of a bound state solution for a free potential equation, and for
this reason, it is the only acceptable one.
For the complete equation, $V(r)\neq 0$,
one must start from this particular extension of the free Hamiltonian
and then add a potential. This does not alter the self-adjointness, provided
the potential is ``weak'' in the sense of Kato. The reason is that as the
system is in the eminence of a bound state, adding
any attractive potential to the Friedrich extension, no matter independently
how weak it could be, provides at least one bound state \cite{Chadan}. The potential to
be considered, must therefore preserve the self-adjointness of the differential
operator, according with the following Kato condition:
\be
\int_{0}^{\infty}r(1+|\ln(r)|)|V(r)|dr<\infty~.\label{kato}
\ee

Provided the interaction potential, $V(r)=-C_sK_0(M_Hr)$, satisfies the Kato condition,
the self-adjointness of the total Hamiltonian is assured.
The Kato condition is also decisive to establish a finite number of bound states
(discrete spectrum) and the semi-boundness of the complete Hamiltonian.
In conclusion, the physical asymptotic solution of
Eq.(\ref{diff3}) is given only by $\sqrt{r}$. In this way the
behavior of the trial function at the origin is completely determined.

On the other hand, at infinity, the trial function must vanish
asymptotically in order to fulfill square integrability. Therefore, a good and suitable
trial function choice (for zero angular momentum) can then be given by
\be
\varphi(r)=\sqrt{r}\exp(-\beta r)~,
\ee
where $\beta$ is a free parameter whose variation approximately
determines an energy minimum.

An analogous procedure can be undertaken to determine
the behavior of a trial function when the angular momentum is different
from zero ($l\neq 0$). In this case, and in the limit $r\rightarrow 0$,
Eq.(\ref{diff2}) reduces to
\be
\left[\frac{\partial^{2}}{\partial r^2}-\frac{l^2-\frac{1}{4}}{r^{2}}+k^{2}\right]u(r)=0~,
\ee
whose general solution reads
\be
u(r)=B_{1}r^{(l+1/2)}+B_{2}r^{(-l+1/2)}~.
\ee
For $l>0,$ the choice $r^{(l+1/2)}$ assures a trial function well-behaved
at the origin. Since the Schr\"odinger equation depends only on $l^2$,
any of the choices, $l>0$ or $l<0$, is enough for
providing the energy values of the physical states and one gets
\be
\varphi(r)=r^{1/2+l}\exp(-\beta r)~,
\ee
where $\beta$ again is a spanning parameter to be numerically fixed
in order to maximize the binding energy.
Though this last result is mathematically correct,
we should point out that the discussion
regarding non-zero angular momentum states here
is merely for the sake of completeness. The true
wave-function in this case should actually stem from the complete
differential equation, for which one should include the angular components
which remain precluded in this approach.
We shall further comment about this question at the end of Section \ref{sec4}.

%%%%%%%%%%%%%%%%%%%%%%%%% SECTION 4 %%%%%%%%%%%%%%%%%%%%%%%%%%%%%%%%%%%%%%
\section{Digression on the high-T$_{\rm c}$ order parameter, pairing
mechanism and effective coupling constant}\label{sec4}
%%%%%%%%%%%%%%%%%%%%%%%%% SECTION 4 %%%%%%%%%%%%%%%%%%%%%%%%%%%%%%%%%%%%%%
Experimental results have revealed that high-T$_{\rm c}$ superconductivity, as
well as the BCS theory, are related to the existence of electron-electron
bound states. Indeed, there are strong evidences that the charge carriers
are pairs of electrons, for instance in the experiments of quantization of
magnetic flux by Gough {\it et al.} \cite{Gough} and the observation of voltage steps
(Shapiro steps) in the current-voltage curves inside Josephson junctions by
Niemeyer {\it et al.} \cite{Gough}. These facts, among others, indicate
that the order parameter of a reliable theory for high-T$_{\rm c}$
superconductivity must consist of a wave-function representing an electron
pair. Now, there arise two fundamental questions: i) the determination of the type of
wave-function-pairing ($s$-wave, as in the case of the usual
BCS  superconductors; $p$-wave, as it is observed in the superfluid state of
$^3$He; or $d$-wave, as in the case of the heavy-fermion superconductors);
ii) the investigation of the physical mechanism underlying the electron-electron
attraction, and its contribution to the effective coupling constant.

Regarding to the first question, i), we should emphasize that
the type of order parameter constitutes a key question for the
understanding of high-T$_{\rm c}$ superconductivity. In the latest
80's,  consensus about the $s$-wave pairing was nearly established due
to some pioneer experiments, {\it e.g.} Josephson tunneling in YBaCuO samples, the
temperature dependence of the penetration depth, $\lambda(T)$, and observation of
persistent supercurrents in rings \cite{Persistent-current}. The early Josephson
experiments \cite{Josephson1} were based on the
conviction that the Josephson tunneling was not feasible between paired
electrons in two different angular momentum states, unless dissipation
occurred in the junction. Experiments with Y$123$ linked to Pb or Sn point
contacts (ordinary BCS superconductors) reported no dissipation, so that Y$123$
was declared to be in $s$-pairing state. Despite the observations of
$\lambda(T)$ indicating an $s$-wave order parameter for some
planar superconductors \cite{Harshman}, experimental verification of a linear
behavior for $\lambda(T)$ was afterwards obtained by Hardy {\it et al.} \cite{Today-Levi},
and theoretically predicted by Annet {\it et al.} \cite{Annett}. Making use of the
ARPES (angle resolved photoemission spectroscopy) technique, Shen {\it et al.} \cite{Shen}
reported on the observation of points of very small gap energy along the diagonal
direction ($|K_x|=|K_y|$) for the BSCCO and YBCO samples,  consistent with a
$d$-wave-function pattern. Other experiments sensitive to phase changes of the order
parameter, composed by DC SQUIDs \cite{DcSQUID}, reiterated the
$d_{x^2-y^2}$-wave-function model.

Nowadays, the status of the situation moved to a
position midway between the two opposing results above discussed.
Recently, a modern interpretation of a
peculiar Josephson tunneling experiment \cite{Josephson-c}, that measures the
tunneling current along the $c$-axis, has shed light on a
new reality concerning the structure of the order parameter.
As a matter of fact, the outcomes obtained by Kouznetsov {\it et al.} \cite{Josephson-c}
in 1997 showed up compatibility only with a {\it mixed} wave, composed by a $d$ plus an
$s$-wave component, as first noticed by Sun {\it et al.} \cite{Josephson-c}.
Indeed, several very recent publications \cite{Mixed-wave}
have claimed on a $s$-wave pattern with admixture of $d$-wave, coming across as a
new area of investigation.
According to some of these studies, it is verified that the $d_{x^2-y^2}$
order parameter is dominant just for the higher temperatures while at lower
ones the order parameter becomes more and more $s$-like, showing up a mixed symmetry.
Actually, the above discussion concerns mainly the order parameter of the usual
high-T$_{\rm c}$ compounds (hole-doped ones). In the case of the electron-doped
cuprates, there are strong experimental evidences \cite{Electron-doped}
supporting the conventional $s$-wave order parameter and suggesting a
BCS-like behavior.

Now, regarding to the second question, ii), in the usual superconductors, the isotope
effect ($T_{\rm c}\sim M^{-\alpha}$, $\alpha=0,5$) was decisive for the establishment
of the BCS-theory, which successfully proposed the lattice vibrations (phonons) to
explain the electron-electron attraction and a symmetric $s$-wave-function
representing the electron-pair. Beyond the scope of the conventional
superconductors, the manifestation of the isotope effect is a rather complex
phenomenon dependent on other factors besides the lattice vibrations, as the
presence of magnetic impurities. In this regard, the deviations from the BCS
reference value ($\alpha\sim 0,5$) observed in many materials, including the
high-T$_{\rm c}$ oxides, cannot be used unequivocally to rule out the phononic
mechanisms from the set of the likely excitations that contribute
effectively to the pairing \cite{Kresin}. Indeed, there exists the general
assumption that the isotope effect and the phonon interaction should be
ubiquitous in the cuprates, but not as the only mechanism yielding the pair
condensation, which leads to the certainty that other mechanisms must coexist
with the phonon one in order to assure the high values of the coupling constant
and the large critical temperatures measured. The nature of these mechanisms
has been an issue of intensive research, and despite the exhaustive efforts
undertaken in this area, no consensus has yet been reached. Among the variety
of approaches to this issue, one can mention some exotic attempts
(non-phonon ones) \cite{Exotic-coupling} pointing to a non-symmetric
solution, as the plasmon-wave excitations \cite{Plasmons}, the magnon interchange
model \cite{Chen-Goddard}, the spin fluctuation interchange model
\cite{Spin-fluctuation}, the excitonic pairing model \cite{Excitons},
the polarons and bipolarons mechanisms \cite{Polarons}.

The fact that the ubiquitous electron-phonon interaction is disguised among
other non-phonon mechanisms, creates an identification problem for
the corresponding coupling constants. In Condensed Matter terminology, the
electron-phonon coupling constant $\lambda_{\rm ep}$ reflects the effect of
collective vibrations (phonons) of the whole lattice on each charge carrier.
Besides the phononic interaction, one considers the existence of other
mechanism, but up to now nobody knows to determine to which extent the
electron-phonon contribution (and the non-phonon ones) participates in the
effective electron-electron interaction. The quantification of the
contributions of each interaction mechanism to the effective attraction
(through the stipulation of values for the coupling constants)
is a question that could be answered only if all the mechanisms were
well-understood.
While the answer is not clear, the option is to work with effective interactions and
coupling constants. In this sense, the coupling constant of interest must be
an effective one, able to account for the contributions of several similar
interactions that, in the case of the present field-theoretic
model, will have a scalar character.

According to the phenomenological picture above described, one should accept the
evidences pointing to a mixed order parameter composed both by $s$-
and $d$-waves in the case of hole-doped cuprates and probably pure $s$
in the case of electron-doped materials.
The present work will deal with the $s$-component in view of the microscopic
field-theoretic scenario we set up.
Our model relies on a mechanism of one-particle exchange (photon and Higgs quasi-particle)
in the non-relativistic limit, to account for the attractive electron-electron potential.
Should we relax the Born approximation and add up loop corrections to the
tree-level amplitudes considered here, an anisotropic potential would come out,
so that it could account also for the $d$-wave contribution as a result
of 1-loop effects (this will be analyzed in a separate paper \cite{d-wave}).
In any case, the present radial (isotropic) potential is entirely
suitable for addressing pure $s$-wave type systems, as it is the case for the
electron-doped high-T$_{\rm c}$ superconductors.

%%%%%%%%%%%%%%%%%%%%%%%%% SECTION 5 %%%%%%%%%%%%%%%%%%%%%%%%%%%%%%%%%%%%%%
\section{Interplay between high-T$_{\rm c}$ phenomenology and planar QED}\label{sec5}
%%%%%%%%%%%%%%%%%%%%%%%%% SECTION 5 %%%%%%%%%%%%%%%%%%%%%%%%%%%%%%%%%%%%%%
The evidence of a quasi-planar structure in high-T$_{\rm c}$ superconductivity
is a suitable reason
for adopting a planar QED model as a
theoretical starting point. However, there arises the necessity of
establishing a relationship between the parameters of the model and the
experimental data for cuprate high-T$_{\rm c}$ superconductors. In the present
parity-preserving action, there are some free parameters that could be
identified with phenomenological observables which are of crucial importance to
describe these materials. In Eq. (\ref{action1}), the electron-Higgs coupling, $y$,
is an effective constant that embodies all possible mechanisms of
interaction between electrons via Higgs-type excitations.
As a result of the scalar
character of this mediation, one encloses a large diversity of spinless
bosonic interaction mechanisms; namely, the phonons,
the plasmons \cite{Plasmons},
and other collective excitations. This theoretical
similarity suggests an
identification of the field theory parameter with an effective
electron-scalar coupling (instead of an electron-phonon one):
$y\rightarrow\lambda_{\rm es}$.
It is expected that the values of $\lambda_{\rm es}$ must be larger than the
values of $\lambda_{\rm ep}$, in view of the effective character of this new coupling
constant, that comprises other interactions besides the phononic case.
It must be said that the
magnetic models based on antiferromagnetic spin-fluctuations (magnons,
spin-polarons, excitons, etc.) support just a $d$-wave order parameter
and suppose an intermediation by 1-spin gauge particles which, if indeed real,
obviously does not contribute to $\lambda_{\rm es}$.

Another well-known and well-measured high-T$_{\rm c}$ superconducting parameter is
related to the magnetic field penetration depth orthogonally to the Cu-O planes,
$\lambda_c$. The observation of an orthogonal parameter in a
quasi-planar system is an indicative inheritance of a third lost
(spatial) dimension. Specifically, in QED$_3$, the electromagnetic coupling
constant squared, $e^2$, has dimension of mass, rather than the dimensionless
character of the usual four-dimensional QED$_4$ coupling constant. This fact might be
understood as a memory (or reminiscence) of the third dimension that appears
(into the coupling constant) when one tries to work with a theory
intrinsically defined in three space-time dimensions.
This dimensional peculiarity could be
better implemented through the definition of a new coupling constant in three
space-time dimensions \cite{Kogan,Randjbar}: $e\rightarrow e_3=e/\sqrt{l}$,
where $l$ represents a distance orthogonal to the planar dimension. This
parameter shall be identified with the $c$-axis magnetic penetration depth
$(\lambda_c)$, whose values will be taken from the phenomenological
data set available for the high-T$_{\rm c}$ cuprate superconductors analyzed here.
In this way, one writes,
$e_3=e/\sqrt{\lambda_c}$,
where from now on the electromagnetic coupling constant, $e$,
is the actual electron charge.
The phenomenological identification of these two parameters will make
the planar Schr\"odinger equation entirely known and, consequently,
will allow the application of a numerical method (such as the variational one)
for computing its energy bound states.

%%%%%%%%%%%%%%%%%%%%%%%%% SECTION 6 %%%%%%%%%%%%%%%%%%%%%%%%%%%%%%%%%%%%%%
\section{\bigskip Numerical evaluation}\label{sec6}
%%%%%%%%%%%%%%%%%%%%%%%%% SECTION 6 %%%%%%%%%%%%%%%%%%%%%%%%%%%%%%%%%%%%%%
In this Section, examples of quasi-planar copper oxide superconductors are
displayed, each one associated to a corresponding energy gap,
($2\Delta(0)$), $c$-axis magnetic penetration depth
($\lambda_c$), electron-scalar coupling ($\lambda_{\rm es}$) and the
corresponding scalar vacuum expectation value squared ($v^2$) that
provides the gap energy; $\beta$ is for the value of the parameter that
minimizes the energy and $C_{s}$ the coefficient of the electron-electron
scattering potential given by Eq.(\ref{potential1}). The numerical procedure is linear;
namely, the choice of
input data ($v^2$, $\lambda_c$, $\lambda_{\rm es}$) determines the coefficient
$C_s=1/2\pi(2\lambda_{\rm es}^2v^2-e^2/\lambda_c)$ and the Higgs quasi-particle
mass $M_H={\sqrt 2}ve/{\sqrt \lambda_c}$ which is the argument of the Bessel
function $K_0$. All this allows the
Schr\"odinger equation (\ref{diff2}) to become totally known. Thereafter, the
application of the variational method allows one to find a value of $\beta$
which provides, up to an uncertainty of $\pm 0,5$meV, the expected gap energy.
The quantity $\xi_{ab}$
represents the average size of the wave-function associated to the computed
bound state, which might be tantamount to the planar
correlation length of the cuprate materials. As a matter of fact,
it can be taken as a suitable measure of the correlation length.

It was already explained that the constant $\lambda_{\rm es}$ constant comprises not only
the phonon contribution, but all the scalar ones. There arises the issue of how
one may estimates the value of this constant.
Regarding $\lambda_{\rm ep}$, one knows that
the experimental techniques brings to light a great
variation of values from a sample to another,
and even for the same sample. For example, in
YBa$_2$Cu$_3$O$_7$ samples, the measurements of $\lambda_{\rm ep}$ vary
from $0,2$ to $2,5$ \cite{Lambda-coupling}, such an indefinite picture occurs
also for other superconductors.
In the case of $\lambda_{\rm es}$, larger values are
expected due to its effective nature, so that in the following
Tables~\ref{table1}-\ref{table4} this
constant will be spanned from $0,5$ to $4,0$. The Tables~\ref{table1}-\ref{table4}
contain data for the zero angular momentum ($l=0$) and singlet-spin state, in order
to account for the $s$-wave pairing structure of superconductors, where the input
data have been collected from the works of Ref.\cite{Maeda-Schilling} for the following
high-T$_{\rm c}$ cuprate superconductors: YBa$_2$Cu$_3$O$_7$,
Tl$_2$Ba$_2$CaCu$_2$O$_{10}$,
Bi$_2$Sr$_2$CaCu$_2$O$_8$ and HgBa$_2$Ca$_2$Cu$_3$O$_8$.
%%%%%%%%%%%%%%%%%%%%%%%%%%%%%%%%%%%%%%%%%%%%%%%%%%%%%%%%%%%%%%%%%%%%%%%%%%%%%%%%%%%%%%%%%%
\begin{table}
%[tbp]
%\par
%\begin{center}
\begin{tabular}{|c|c|c|c|c|c|c|}
$v^2$(meV) & $\lambda_c$(\AA) & $\lambda_{\rm es}$
& $C_s$(meV) & $\beta$ & $E_{\rm gap}\pm 0,5$(meV) & $\xi_{ab}$(\AA)\\
\hline\hline
71,33 & 1800 & 0,5 & 4,40 & 33,52 & 29,9 & 29,43\\\hline
16,65 & 1800 & 1,0 & 4,03 & 32,07 & 30,1 & 30,76\\\hline
7,10 & 1800 & 1,5 & 3,82 & 31,21 & 30,0 & 31,61\\\hline
3,90 & 1800 & 2,0 & 3,69 & 30,71 & 30,1 & 32,13\\\hline
2,44 & 1800 & 2,5 & 3,58 & 30,23 & 30,0 & 32,64\\\hline
1,67 & 1800 & 3,0 & 3,51 & 29,97 & 30,0 & 32,92\\\hline
1,22 & 1800 & 3,5 & 3,48 & 29,84 & 30,3 & 33,06\\\hline
0,92 & 1800 & 4,0 & 3,41 & 29,52 & 30,2 & 33,42\\
%\hline
\end{tabular}
%\end{center}
\caption{Input (from Hasegawa {\it et al.} and Gallagher {\it et al.}
[44]) and output data for YBa$_2$Cu$_3$O$_7$
(T$_{\rm c}=87$K and $2\Delta(0)=30,0$meV).}\label{table1}
\end{table}
%%%%%%%%%%%%%%%%%%%%%%%%%%%%%%%%%%%%%%%%%%%%%%%%%%%%%%%%%%%%%%%%%%%%%%%%%%%%%%%%%%%%%%%%%%
\begin{table}
%[tbp]
%\par
%\begin{center}
\begin{tabular}{|c|c|c|c|c|c|c|}
$v^2$(meV) & $\lambda_c$(\AA) & $\lambda_{\rm es}$
& $C_s$(meV) &
$\beta$ & $E_{\rm gap}\pm 0,5$(meV) & $\xi_{ab}$(\AA)\\
\hline\hline
54,00 & 4800 & 0,5 & 3,82 & 31,23 & 28,1 & 31.59\\\hline
12,50 & 4800 & 1,0 & 3,50 & 29,94 & 28,1 & 32,95\\\hline
5,30 & 4800 & 1,5 & 3,31 & 29,12 & 28,0 & 33,88\\\hline
3,10 & 4800 & 2,0 & 3,29 & 29,03 & 28,1 & 33,99\\\hline
1,92 & 4800 & 2,5 & 3,17 & 28,42 & 27,7 & 34,72\\\hline
1,32 & 4800 & 3,0 & 3,13 & 28,28 & 28,0 & 34,89\\\hline
0,95 & 4800 & 3,5 & 3,05 & 27,92 & 27,7 & 35.34\\\hline
0,72 & 4800 & 4,0 & 3,01 & 27,73 & 27.8 & 35,58\\
%\hline
\end{tabular}
%\end{center}
\caption{Input (from Hasegawa {\it et al.} and Thompson {\it et al.}
[44]) and output data for Tl$_2$Ba$_2$CaCu$_2$O$_{10}$
(T$_{\rm c}=105$K and $2\Delta(0)=28,0$meV).}\label{table2}
\end{table}
%%%%%%%%%%%%%%%%%%%%%%%%%%%%%%%%%%%%%%%%%%%%%%%%%%%%%%%%%%%%%%%%%%%%%%%%%%%%%%%%%%%%%%%%%%
\begin{table}
%[tbp]
%\par
%\begin{center}
\begin{tabular}{|c|c|c|c|c|c|c|}
$v^2$(meV) & $\lambda_c$(\AA) & $\lambda_{\rm es}$ & $C_s$(meV) & $\beta$
& $E_{\rm gap}\pm 0,5$(meV) & $\xi_{ab}$(\AA)\\
\hline\hline
96,5 & 5000 & 0,5 & 7,22 & 43,02 & 53,4 & 22,93\\\hline
22,2 & 5000 & 1,0 & 6,61 & 41,11 & 53,4 & 23,99\\\hline
9,42 & 5000 & 1,5 & 6,29 & 40,11 & 53,4 & 24,59\\\hline
5,14 & 5000 & 2,0 & 6,09 & 39,41 & 53,4 & 25,04\\\hline
3,21 & 5000 & 2,5 & 5,93 & 38,95 & 53,3 & 25,33\\\hline
2,19 & 5000 & 3,0 & 5,82 & 38,59 & 53,4 & 25,57\\\hline
1,59 & 5000 & 3,5 & 5,72 & 38,22 & 53,4 & 25,81\\\hline
1,20 & 5000 & 4,0 & 5,65 & 38,06 & 53,5 & 25,92\\
%\hline
\end{tabular}
%\end{center}
\caption{Input (from Maeda [44]) and output data
for Bi$_2$Sr$_2$CaCu$_2$O$_8$ (T$_{\rm c}=109$K and $2\Delta(0)=53,4$meV).}\label{table3}
\end{table}
%%%%%%%%%%%%%%%%%%%%%%%%%%%%%%%%%%%%%%%%%%%%%%%%%%%%%%%%%%%%%%%%%%%%%%%%%%%%%%%%%%%%%%%%%%
\begin{table}
%[tbp]
%\par
%\begin{center}
\begin{tabular}{|c|c|c|c|c|c|c|}
$v^2$(meV) & $\lambda_c$(\AA) & $\lambda_{\rm es}$ & $C_s$(meV) & $\beta$
& $E_{\rm gap}\pm 0,5$(meV)
& $\xi_{ab}$(\AA)\\
\hline\hline
92,00 & 3500 & 0,5 & 6,67 & 41,28 & 48,0 & 23,90\\\hline
21,20 & 3500 & 1,0 & 6,09 & 39,48 & 48,1 & 24,99\\\hline
9,00 & 3500 & 1,5 & 5,79 & 38,49 & 48,0 & 25,63\\\hline
4,90 & 3500 & 2,0 & 5,58 & 37,78 & 47,9 & 26,12\\\hline
3,07 & 3500 & 2,5 & 5,45 & 37,31 & 48,0 & 26,44\\\hline
2,10 & 3500 & 3,0 & 5,36 & 37,03 & 48,2 & 26,64\\\hline
1,51 & 3500 & 3,5 & 5,25 & 36,61 & 47,9 & 26,95\\\hline
1,15 & 3500 & 4,0 & 5,19 & 36,41 & 48,1 & 27,09\\
%\hline
\end{tabular}
%\end{center}
\caption{Input (from Schilling {\it et al.} [44])
and output data for HgBa$_2$Ca$_2$Cu$_3$O$_8$
(T$_{\rm c}=131$K and $2\Delta(0)=48,0$meV).}\label{table4}
\end{table}
%%%%%%%%%%%%%%%%%%%%%%%%%%%%%%%%%%%%%%%%%%%%%%%%%%%%%%%%%%%%%%%%%%%%%%%%%%%%%%%%%%%%%%%%%%
%%%%%%%%%%%%%%%%%%%%%%%%% SECTION 7 %%%%%%%%%%%%%%%%%%%%%%%%%%%%%%%%%%%%%%
\section{Final remarks}
%%%%%%%%%%%%%%%%%%%%%%%%% SECTION 7 %%%%%%%%%%%%%%%%%%%%%%%%%%%%%%%%%%%%%%
Starting off from a parity-preserving planar QED model
\cite{phdthesis,N.Cimento,Del_cima},
this paper sets out to mainly evaluate the energy of the ground state of
electron-pairs that interact via photon and Higgs quasi-particle exchange.
The numerical results,
obtained throughout a variational method, succeeded in fitting
some well-known
parameters, such as energy gap, for the high-T$_{\rm c}$ copper oxide superconductors
analyzed here, namely, YBa$_2$Cu$_3$O$_7$,
Tl$_2$Ba$_2$CaCu$_2$O$_{10}$,
Bi$_2$Sr$_2$CaCu$_2$O$_8$ and HgBa$_2$Ca$_2$Cu$_3$O$_8$.
One has therefore a theoretical model which,
supplemented by some
experimental data on high-T$_{\rm c}$ superconductors, may reveal itself
suitable for treating quasi-planar superconductivity.
An important outcome is that the phenomenological
data fix the scale for the breaking of the $U(1)$-symmetry in the
superconductors: $v^2\sim 1$-$10$meV, in much the same way as $\sim10^2$GeV
is the scale for the breaking of the electroweak symmetry
in the Standard Model. In the
described picture, the Higgs mechanism plays an essential role in providing
mass for the photons and to ensure a net
attractive electron-electron scattering potential through the exchange of photons
and Higgs quasi-particles.

The potential resulting from the M{\o}ller scattering in the non-relativistic
limit, $-C_sK_0(M_Hr)$,
provides just a symmetric wave-function solution to the order parameter.
The search
for an anisotropic wave-function ($p$-wave or $d$-wave) must pass through the
attainment of a potential dependent on the angle variable,
in such a way that
it may account for the angular variations observed in these
non-symmetric states. We will hopefully arrive at the angular dependence by including
loop corrections into the scattering potential \cite{d-wave}.

\section*{Acknowledgements}
The authors would like to thank Prof. Yu Lu and Prof. Z. Te\v{s}anovi\'c for having
pointed out some relevant references. M.M.F.Jr. would like to thank the
HE-Section of the Abdus Salam International Centre for Theoretical Physics (ICTP)
for the kind hospitality and financial support, and to its Head, Prof. S. Randjbar-Daemi.
Thanks are also due to the Head of CFC-CBPF,
Prof. A.O. Caride, for the financial support and encouragement.
O.M.D.C. dedicates this work to his wife, Zilda Cristina,
to his kids, Vittoria and Enzo, and his mother, Victoria.
H.R.C. is grateful to Maro Tinto Osses and Enano Iriarte for stimulating and clarifying
remarks, and to Alela Santillan Iturr\'es and Mayo Nesa Gioia for their friendship and
comments during his stay at La Plata, Argentina;  Baronesa Silvia 
and {\it Le Baron} Rousseau Portalis
are gratefully acknowledged for their kind hospitality in Buenos Aires, and Fidel Arturo 
Schaposnik, head of LFT, 
Departamento de F\'\i sica de La Plata, for financial support. H.R.C dedicates
this work to  Lili, Gaby, Agustin, Clarita, M\'aximo and Juan Cruz.

\end{document}